\documentclass[11pt]{article}
      
\usepackage{a4wide}
\usepackage{amsmath}
\usepackage{amssymb}           
\usepackage{exscale}
\usepackage{cite}

\newcommand{\mZ}{\mathbb{Z}}
\newcommand{\mN}{\mathbb{N}}

\newcommand{\Gp}[1]{G^+_{#1}}
\newcommand{\Gm}[1]{G^-_{#1}}

\newcommand{\ttau}{\tilde{\tau}}
\newcommand{\qt}{\tilde{q}}

\newcommand{\ket}[1]{|#1\rangle}

\newcommand{\Iket}[1]{|#1\rangle\!\rangle}
\newcommand{\Ibra}[1]{\langle\!\langle#1|}
\newcommand{\Bket}[1]{|\!|#1\rangle\!\rangle}
\newcommand{\Bbra}[1]{\langle\!\langle#1|\!|}
\newcommand{\bZ}{\bar{Z}}
\newcommand{\Zbar}{\bZ}
\renewcommand{\d}{\partial}

\newcommand{\alphabar}{\bar{\alpha}}

\DeclareMathOperator{\mTr}{Tr}

\newcommand{\vt}[2]{\vartheta_{#1}(#2)}
\newcommand{\vth}{$\vartheta$}

\renewcommand{\paragraph}[1]{
  \vspace{0.5cm}
  \pagebreak[1]
  \noindent
  {\it #1}
  \nopagebreak
  \vspace{0.3cm}
  \nopagebreak}

\numberwithin{equation}{section}

\begin{document}

\begin{titlepage}
\renewcommand{\thefootnote}{\fnsymbol{footnote}}
  \flushright{KCL-MTH-04-04 \\
              hep-th/0404062}
  
  \begin{center}
    
    \vspace{2cm}
    {\LARGE \textbf{$N=2$ Superconformal boundary states for free}}
\vspace{0.5cm}

{\LARGE \textbf{bosons and fermions}} 
    
    \vspace{2cm}
     {\large Matthias R. Gaberdiel}\footnote{\tt{mrg@phys.ethz.ch}}\\    
    \vspace{0.3mm}
    {\it Institut f{\"u}r Theoretische Physik\\ ETH Z{\"u}rich \\
    CH-8093 Z{\"u}rich\\ Switzerland \\ }
    \vspace{.5cm}     
    and  

    \vspace{.5cm} 
    {\large Hanno Klemm}\footnote{\tt{klemm@phys.ethz.ch}}\\    
    \vspace{0.3mm}
    {\it Department of Mathematics \\
    King's College London \\
    Strand\\
    London WC2R 2LS \\ UK \\
    \vspace{0.3mm}
    and \\
    \vspace{0.3mm}
    Institut f{\"u}r Theoretische Physik\\ ETH Z{\"u}rich \\
    CH-8093 Z{\"u}rich \\ Switzerland }

    \vspace{1cm}
\begin{abstract}
The most general $N=2$ superconformal boundary states for the $c=3$
theory consisting of two (uncompactified) free bosons and fermions are
constructed. It is shown that the only $N=2$ boundary states are the
familiar Dirichlet boundary states, as well as the Neumann boundary
states with an arbitrary electric field.
\end{abstract}

\end{center}
\end{titlepage}

\section{Introduction}
\label{sec:introduction}
\setcounter{footnote}{0}

Most D-branes that have been considered in string theory so far are
rather special in the sense that they preserve more symmetries than is
actually required for their consistency. For example, for the case of
string theory in flat space, the usual Neumann and Dirichlet branes 
preserve all the U(1) current symmetries associated to the spacetime
coordinates. In terms of their boundary states, this means that they
satisfy the gluing conditions 
\begin{equation}
  \label{eq:75}
  (\alpha_n^i \pm \tilde{\alpha}^i_{-n})\Bket{B} = 0\,,
\end{equation}
where the $+ (-)$ sign applies if $i$ is a Neumann (Dirichlet)
direction. However, as was already pointed out by Polchinski
\cite{Polchinski:1994fq}, the consistency of the construction only
requires that the boundary state preserves the conformal symmetry  
\begin{equation}
  \label{eq:76}
  (L_n - \tilde{L}_{-n})\Bket{B}=0\,.
\end{equation}
Since $L_n$ is quadratic in the $\alpha$'s, equation \eqref{eq:75}
implies equation \eqref{eq:76} but the converse is not true. It is
therefore interesting to determine the additional boundary states 
that only preserve the conformal symmetry \eqref{eq:76}, but not
\eqref{eq:75}. 

For the case of the theory consisting of a single free boson, this
question was answered in 
\cite{Friedan,Gaberdiel:2001xm,Gaberdiel:2001zq,Janik:2001hb,Tseng:2002ax},
and the 
generalisation to the $N=1$ supersymmetric theory consisting of a free
boson and a 
free fermion was considered in \cite{Gaberdiel:2001zq}. While there
are, in both cases, new D-branes other than the usual Neumann
and Dirichlet branes, none of the additional branes are stable. 
In fact none of these branes preserves spacetime supersymmetry, since 
spacetime supersymmetry requires an unbroken $N=2$ worldsheet
supersymmetry \cite{Banks:1988cy}, which was not present in those
theories. It is therefore of particular importance to study the above
question in the context of an $N=2$ worldsheet theory. The simplest
example of such a theory is the theory of two free bosons and fermions
with $c=3$, and this is the theory we shall be analysing in this
paper. As we shall show, in this case all $N=2$ superconformal
D-branes can be accounted for in terms of the usual Neumann and
Dirichlet branes. This is quite a surprising result since one may have
thought by analogy with the situation for $N=1$ supersymmetry that
there should be many more $N=2$ D-branes. On the other hand, our
result is quite reassuring since it implies that, at least in this
example, the usual Dirichlet and Neumann branes already account for
all spacetime supersymmetric branes. One expects that there are other
backgrounds for which one should be able to find `new' $N=2$ D-branes;
we believe that the techniques that are described in this paper will
help to construct these D-branes.

D-branes that preserve at least the $N=2$ supersymmetry have been
studied before in the framework of Gepner models and Calabi-Yau
compactifications (see for example
\cite{Ooguri:1996ck,Recknagel:1998sb}), for $N=2$ Landau-Ginzburg
models \cite{HIV} and non-linear sigma models (see for example
\cite{HIV,Lindstrom:2002jb,ALZ}) and for $N=2$ minimal models and
coset models \cite{MMS,FS,Parkhomenko:2003gy,Stefan}. More recently
$N=2$ D-branes have also been analysed in Liouville theory
\cite{Eguchi:2003ik,Ahn1,Ahn2}. In most cases however, the D-branes that have
been considered preserve more symmetry than just the $N=2$
algebra. Obviously, for $N=2$ minimal models all $N=2$ D-branes are
known, but this is simply a consequence of the fact that the whole
theory is rational with respect to this algebra; in our example, the
theory is not rational with respect to the $N=2$ algebra, and it is
therefore a priori not obvious what the most general $N=2$ D-branes
should be. 
\smallskip

The paper is organised as follows. In section \ref{sec:setup} we 
review some relevant facts about the $N=2$ superconformal algebra and
its free field realisation. The Neveu-Schwarz boundary states are
constructed in section \ref{sec:boundary-states}, and in 
section \ref{sec:amplitude} we consider the cylinder amplitude
of two such boundary states. These amplitudes agree with the
amplitudes of the usual Neumann boundary states with electric flux, as
follows from some non-trivial \vth-function identity which we derive
from first principles. Furthermore we discuss the GSO projection for
the NS boundary states. In section \ref{sec:ramond-sector} the
analysis is repeated for the Ramond sector and in section
\ref{sec:normalisation} we fix the normalisation constants of the
boundary states by comparison with the open string amplitude. Chapter
\ref{sec:conclusion} contains our conclusions. There are three
appendices that describe some of the more technical calculations.

\section{The $N=2$ algebra and its free field realisation}
\label{sec:setup}

Let us begin by reviewing the $N=2$ superconformal algebra, and its
realisation in terms of two free bosons and fermions.

The $N=2$ algebra consists of the energy momentum tensor, two
supercharges $G^\pm$ and a U(1) current $J$. The commutation relations
are given by \cite{Ademollo:1976an}
\begin{align*}
  \left[L_m, L_n \right] &=(m-n)L_{m+n} + \tfrac{c}{12}(m^3 -
  m)\delta_{m,-n} \,, \\ 
  \left[L_m, J_n \right] &=-nJ_{m+n}\,,\\
  \left[L_m, G_r^\pm \right] &=\left( \tfrac{1}{2}m-r \right)
  G_{m+r}^\pm \,, \\
  \left[J_m, J_n \right] &= \tfrac{c}{3} m\delta_{m,-n}\,,\\
  \left[J_m, G_r^\pm \right] &=\pm G_{r+m}^\pm \,, \\
  \left\{ \Gp{r}, \Gm{s}\right\} &= 2 L_{r + s} + (r -s) J_{r + s} +
  \tfrac{c}{3}(r^2 - \tfrac{1}{4})\delta_{r, -s} \,, \\
  \left\{\Gp{r}, \Gp{s} \right\} &=\left\{\Gm{r}, \Gm{s}\right\}=0\,,
\end{align*}
where $m,n$ are integers and $r,s$ are half-integers (integers) in the
Neveu-Schwarz (Ramond) sector. For the free field realisation of the
algebra in terms of two free bosons and fermions the 
central charge equals $c=3$. We will denote the eigenvalue  of
$L_0$ (the \emph{weight}) by $h$ and the eigenvalue of $J_0$ (the
\emph{charge}) by $Q$. 

In order to represent this algebra in terms of free fields we will
arrange the two free bosonic fields $X^1(z,\bar{z})$, $X^2(z,\bar{z})$
and the two free real fermions $\psi^1(z)$, $\psi^2(z)$ in complex
pairs 
\begin{align*}
  Z &=\frac{1}{\sqrt{2}}(X^1 + iX^2)\,, & 
  \bZ &= \frac{1}{\sqrt{2}}(X^1 - iX^2)\,, \\  
  \psi^+&=\frac{1}{\sqrt{2}}(\psi^1 + i \psi^2)\,, & 
  \psi^- &=\frac{1}{\sqrt{2}}(\psi^1 - i \psi^2)
\end{align*}
with mode expansions (for the left-movers)
\begin{subequations}
  \label{eq:84}
\begin{align}
  \d Z &= -i \sum_{m=-\infty}^{\infty}\alpha_mz^{-m-1}\,, &
  \d \Zbar &= - i \sum_{m=-\infty}^{\infty}\alphabar_m
  z^{-m-1}\,, \\
  \psi^+ &= \sum_{r} \psi^+_r z^{-r-1/2}\,, &
  \psi^- &= \sum_{r} \psi^-_r z^{-r-1/2}\,.
\end{align}
\end{subequations}
These modes then have the commutation relations
\begin{subequations}
  \label{eq:87}
  \begin{align}  
  [\alpha_m, \alphabar_n] &= m\delta_{m,-n}\,, \\ 
  [\alpha_m,\alpha_n] &= [\alphabar_m, \alphabar_n] = 0\,, \\
  \{\psi^+_r,\psi^-_s\} &= \delta_{r,-s}\,, \\
  \{\psi^+_r,\psi^+_s\}&=\{\psi^-_r,\psi^-_s\}=0\,.\label{eq:96} 
\end{align}
\end{subequations}
Obviously, similar commutation relations hold for the corresponding
right-moving degrees of freedom (that we shall denote by tildes in the
following). 
 
\noindent In terms of the free fields the left-moving $N=2$ algebra is
given by  
\begin{subequations}
  \label{eq:10}
  \begin{align}
    T &=-\d Z \d \bZ -\frac{1}{2}\left( \psi^-\d \psi^+ + \psi^+\d
      \psi^- \right)\,, \\
    G^+ &= i\sqrt{2}\psi^+ \d \bZ\,, \\
    G^- &= i\sqrt{2}\psi^- \d Z\,, \\
    J &= -\psi^{-}\psi^{+}\,,
  \end{align}  
\end{subequations}
and we thus have the mode expansion for the left-moving $N=2$ generators
\begin{subequations}
  \label{eq:89}
  \begin{align}      
    L_n &= \sum_{m} :\alpha_{n-m}\alphabar_{m}: +
    \frac{1}{2}\sum_{s}(2s-n):\psi^{-}_{n-s}\psi^+_s: \,,\\
    J_n &= -\sum_{s}:\psi^-_{-s}\psi^+_{s+n}: \,,\\
    G^+_r &= \sqrt{2}\sum_m :\alphabar_m\psi^+_{r-m}: \,,\\
    G^-_r &= \sqrt{2}\sum_m :\alpha_m\psi^-_{r-m}: \,.
  \end{align}  
\end{subequations}
Obviously, the formulae for the right-moving generators are
identical. 
\medskip

The irreducible representations of the free boson and fermion theory
are labelled by the momenta of the ground state, 
$({\bf p}_L,{\bf p}_R)$, where ${\bf p}_L$ and ${\bf p}_R$ are
two-dimensional vectors. In general the momenta 
$({\bf p}_L,{\bf p}_R)$ lie on a lattice (that characterises the
structure of the two-dimensional torus on which the theory is
compactified); the results depend then in a somewhat 
complicated manner on the precise nature of this lattice. In order to
avoid this difficulty we shall consider here the case where the theory
is uncompactified. Then ${\bf p}:= {\bf p}_L={\bf p}_R$, and 
${\bf p}$ can take any value. The conformal weight of the momentum
ground state labelled by ${\bf p}$ is  
$h=\tilde{h}= \tfrac{1}{2}(p_1^2 + p_2^2)$, where 
$\mathbf{p}=(p_1,p_2)$.

The first step in constructing the most general $N=2$ boundary states
for this theory consists of determining all $N=2$ Ishibashi states. To
this end, we need to decompose the left- and right-moving
representations of the free bosons and fermions in terms of $N=2$
representations. In the uncompactified theory, the left- and
right-moving representations are obviously isomorphic, and it is
therefore sufficient to concentrate just on the left-movers, say. We
shall consider the NS and the R sector separately. 

\subsection{The NS sector}

In the NS sector, it follows from \eqref{eq:89} that each momentum
ground state has U(1) charge $Q=0$. All $N=2$ highest weight
representations for which the highest weight state has $Q=0$ are
irreducible, unless $h=0$, {\it i.e.}\ unless the representation is
the vacuum representation \cite{doerrzapf}. Thus for each 
${\bf p}\ne 0$, the corresponding irreducible free field
representation actually defines an irreducible representation of the
$N=2$ algebra.

The situation is more interesting for the vacuum representation since
the free field representation is then reducible with respect to the
action of the $N=2$ algebra. In fact, the free field representation
contains the $N=2$ singular vectors
\begin{subequations}
  \label{eq:17}
\begin{align}
  v^{(+)}_{(n)} &= (\alpha_{-1})^n\psi^{+}_{-1/2}\ket{0} \,,\\
  v^{(-)}_{(n)} &= (\alphabar_{-1})^n\psi^{-}_{-1/2}\ket{0}\,. 
\label{eq:36}
\end{align}
\end{subequations}
It is shown in appendix \ref{sec:singular-vectors} that these vectors
are actual singular vectors with respect to the $N=2$ algebra. In
fact, these are the only singular $N=2$ vectors, and we therefore have
the decomposition   
\begin{equation}
\label{eq:3}
\mathcal{H}_0^{\text{NS,free}} = \mathcal{H}_{(0,0)}^{\text{NS},N=2}\,
\bigoplus_{n=0}^{\infty}
\left(\mathcal{H}_{(h=n+\frac{1}{2},Q=1)}^{\text{NS},N=2}\oplus
\mathcal{H}_{(h=n+\frac{1}{2},Q=-1)}^{\text{NS},N=2}\right) \,,
\end{equation}
where $\mathcal{H}^{\text{NS},N=2}_{(h,Q)}$ is the irreducible $N=2$
NS-representation with highest weight $h$ and charge $Q$. In order to
see this, one observes that the character of the free field
representation for any $\mathbf{p}$ is given by
\begin{equation}
  \label{eq:2}
\chi_{\mathbf{p}}^{\text{NS,free}}(q)=\mTr{q^{2(L_0-\frac{c}{24})}} =
  q^{\mathbf{p}^2-\frac{1}{4}}\prod_{n=1}^\infty \frac{ 
  (1+q^{2n-1})^2}{(1-q^{2n})^2} \,,
\end{equation}
where $q=e^{i \pi \tau}$. On the other hand, the characters of the
$N=2, c=3$ irreducible highest weight representations are
\cite{Klemm:2003vn}  
\begin{subequations}
  \label{eq:4}
\begin{align}
  \chi^{\text{NS},N=2}_{(0,0)}(q) &= 
  \phi_A(q)\left(1- \frac{2q}{1+q}\right) \,, \\
  \chi_{(h=\frac{2n+1}{2},Q=\pm1)}^{\text{NS},N=2}(q) &= q^{2h}
  \phi_A(q)\left( 1 - q^2 - \frac{q^{2h}}{ 1 + q^{2h}} + \frac{q^{2h+4}}{1
  + q^{2h+2}} \right)\,,
\end{align}
\end{subequations}
where $\phi_A$ is defined as 
\begin{equation}
  \label{eq:5}
  \phi_A(q):=q^{-\frac{1}{4}}\prod_{n=1}^\infty \frac{
  (1+q^{2n-1})^2}{(1-q^{2n})^2}\,.
\end{equation}
It is thus easy to see that
\begin{equation}
\label{eq:6}
\chi_{0}^{\text{NS,free}}(q) = \chi_{(0,0)}^{\text{NS},N=2}(q) +
\sum_{n=0}^{\infty} \left[ \chi_{(\frac{2n+1}{2},1)}^{\text{NS},N=2}(q) +
\chi_{(\frac{2n+1}{2},-1)}^{\text{NS},N=2}(q) \right]\,,  
\end{equation}
which therefore implies that the free field representation cannot
contain any further $N=2$ representations.

\subsection{The R sector}

In the Ramond sector the situation is similar. Any free field
Ramond representation based on ${\bf p}\ne 0$ defines an irreducible
$N=2$ representation. On the other hand, the vacuum representation is
decomposable with respect to the $N=2$ algebra. The actual
decomposition can be obtained from the result in the NS sector using
the spectral flow \cite{SchS,Klemm:2003vn} 
\begin{equation}
\label{eq:3R}
\mathcal{H}_0^{\text{R,free}} = 
\mathcal{H}_{(\frac{1}{8},\frac{1}{2})}^{\text{R},N=2}\, \oplus 
\mathcal{H}_{(\frac{1}{8},-\frac{1}{2})}^{\text{R},N=2}\,
\bigoplus_{n=1}^{\infty}
\left(\mathcal{H}_{(n+\frac{1}{8},\frac{3}{2})}^{\text{R},N=2}\oplus
\mathcal{H}_{(n+\frac{1}{8},-\frac{3}{2})}^{\text{R},N=2}\right) \,,
\end{equation}
where $\mathcal{H}^{\text{R},N=2}_{(h,Q)}$ is the irreducible $N=2$
R-representation with highest weight $h$ and charge $Q$. We should
note that the two Ramond ground states 
$\ket{\tfrac{1}{8}, \pm \tfrac{1}{2}}$ are not related to one another
by the $N=2$ algebra, 
\begin{equation}
  \label{eq:1}
  G^{\pm}_0\ket{\tfrac{1}{8},\tfrac{1}{2}}
=G^{\pm}_0\ket{\tfrac{1}{8},-\tfrac{1}{2}} = 0\,,
\end{equation}
but appear both in the free field representation since 
\begin{equation}
  \label{eq:8}
  \psi^\mp_0\ket{\tfrac{1}{8}, \pm \tfrac{1}{2}} =
  \ket{\tfrac{1}{8},\mp\tfrac{1}{2}}\,.
\end{equation}
They give therefore rise to two separate $N=2$ representations. The
other $N=2$ representations that appear are generated by the singular
vectors 
\begin{subequations}
  \label{eq:60}
\begin{align}
  w^{(+,+)}_{(n)} &=
  (\alpha_{-1})^{n-1}\psi^{+}_{-1}\ket{\tfrac{1}{8},\tfrac{1}{2}} \,,\\ 
  w^{(+,0)}_{(n)} &= (\alpha_{-1})^{n}\ket{\tfrac{1}{8},\tfrac{1}{2}}=
  \Gm{0}w^{(+,+)}_{(n)}
\end{align}
\end{subequations}
and
\begin{subequations}
  \label{eq:61}
\begin{align}
  w^{(-,-)}_{(n)} &=
  (\alphabar_{-1})^{n-1}\psi^{-}_{-1}\ket{\tfrac{1}{8},-\tfrac{1}{2}}\,,
  \\ 
  w^{(-,0)}_{(n)} &=
  (\alphabar_{-1})^{n}\ket{\tfrac{1}{8},-\tfrac{1}{2}} =
  \Gp{0}w^{(-,-)}_{(n)}\,,
\end{align}
\end{subequations}
where $n\in\mN$. The first two singular vectors $w^{(+,+)}_{(n)}$ and
$w^{(+,0)}_{(n)}$ (that are related to one another by the action of
$G^\pm_0$) generate the R-representation  
$\mathcal{H}_{(n+\frac{1}{8},\frac{3}{2})}^{\text{R},N=2}$, while the two
singular vectors $w^{(-,-)}_{(n)}$ and $w^{(-,0)}_{(n)}$ generate 
$\mathcal{H}_{(n+\frac{1}{8},-\frac{3}{2})}^{\text{R},N=2}$. 

Again \eqref{eq:3R} is consistent with the corresponding character
formulae. For the free field representation we simply have 
\begin{equation}
\chi_0^{\text{R,free}}(q) = 2\prod_{n=1}^\infty
  \frac{(1+q^{2n})^2}{(1-q^{2n})^2}\,,
\end{equation}
while the $N=2$ characters are 
\begin{subequations}  
  \label{eq:7}
\begin{align}
\chi_{(\frac{1}{8},\pm\frac{1}{2})}^{\text{R},N=2}(q) &= 
\phi_P(q)\left(1-\frac{q^2}{1+q^2} -\frac{1}{2}\right) \,,\\
\chi_{(n+\frac{1}{8},\pm\frac{3}{2})}^{\text{R},N=2}(q) &= q^{2n}  \phi_P(q)
\left(1-q^2-\frac{q^{2n}}{1+q^{2n}} + \frac{q^{2n+4}}{1+q^{2n+2}}\right)
\end{align}
\end{subequations}
with 
$$
\phi_P(q):=2\prod_{n=1}^\infty\frac{(1+q^{2n})^2}{(1-q^{2n})^2}\,.
$$
As in the NS case it is then obvious that 
\begin{equation}
\label{eq:90}
\chi_0^{\text{R,free}}(q) = 
\chi^{\text{R}, N=2}_{(\frac{1}{8},\frac{1}{2})}(q) +
\chi^{\text{R}, N=2}_{(\frac{1}{8},\frac{-1}{2})}(q) +
\sum_{n=1}^\infty\left(
\chi_{(n+\frac{1}{8},\frac{3}{2})}^{\text{R},N=2}(q) 
+ \chi_{(n+\frac{1}{8},-\frac{3}{2})}^{\text{R},N=2}(q)\right)\,.
\end{equation}
Note that in order to distinguish representations that only differ by
their U(1) charge it would be necessary to include in the trace the
operator $z^{J_0}$; however, the specialised characters used above are
already sufficient for the construction of the boundary states that
shall be considered in the following.

\section{Construction of the boundary states}
\label{sec:boundary-states}

Now we want to construct the boundary states which preserve the
$N=2$ superconformal algebra. There are two possible gluing conditions
which can be imposed on the boundary. They are usually referred to as
A-type and B-type boundary conditions \cite{Ooguri:1996ck}. In terms of
boundary conformal field theory, B-type boundary conditions
correspond to the identity gluing automorphism, whereas A-type boundary
conditions involve the mirror automorphism 
\cite{Lerche:1989uy}. More specifically, the A-type gluing
conditions read 
\begin{subequations}
  \label{eq:11}
\begin{align}
  (L_n - \tilde{L}_{-n})\Bket{B}_A&=0 \,,\\
  (J_n - \tilde{J}_{-n})\Bket{B}_A&=0 \,,\\
  (\Gp{r} + i\,\eta\, \tilde{G}^-_{-r})\Bket{B}_A&=0 \,,\\
  (\Gm{r} + i\,\eta\, \tilde{G}^+_{-r})\Bket{B}_A&=0 \,,
\end{align}
\end{subequations}
while the B-type gluing conditions are 
\begin{subequations}
  \label{eq:12}
  \begin{align}
  (L_n - \tilde{L}_{-n})\Bket{B}_B&=0 \,,\\
  (J_n + \tilde{J}_{-n})\Bket{B}_B&=0 \,,\\
  (\Gp{r} + i\,\eta\, \tilde{G}^+_{-r})\Bket{B}_B&=0 \,,\\
  (\Gm{r} + i\,\eta\, \tilde{G}^-_{-r})\Bket{B}_B&=0\,.
\end{align}
\end{subequations}
Here $\eta=\pm$ describes the different spin structures. For the free
field situation considered here, A-type and B-type gluing conditions
are essentially equivalent (since `mirror symmetry' corresponds 
simply to T-duality in one direction); we shall therefore concentrate
on the technically simpler B-type case in the following.

These gluing conditions have only non-trivial solutions for the NS-NS
and the R-R sector. In this section we want to consider the component
of the boundary state in the NS-NS sector; the construction 
for the R-R sector (that is fairly similar) will be done in section 
\ref{sec:ramond-sector}.

\subsection{The boundary states in the NS-NS sector}

In the first step we need to identify all the $N=2$ Ishibashi states
that arise in the NS-NS sector. Given that the left- and right-moving
representations are isomorphic (as free field representations), and
that we consider the trivial gluing condition, it follows from the
standard arguments that there is precisely one $N=2$ Ishibashi state
for each irreducible $N=2$ representation that appears in the above 
decompositions \cite{Ishibashi:1989kg}. In particular, we therefore
have one $N=2$ Ishibashi state for each ${\bf p}\ne 0$; this Ishibashi
state must therefore agree with the usual U(1) Dirichlet Ishibashi
state. By considering the usual Dirichlet branes, we can construct
boundary states that couple to all of these Ishibashi states, and they
therefore do not give rise to any new $N=2$ boundary states. 

The situation is more interesting for the vacuum sector. For the U(1)
theory, there is one Ishibashi state (for each gluing condition) that
can be associated to this sector; more specifically, we have
one U(1) Dirichlet and one U(1) Neumann Ishibashi state (for each
choice of $\eta$) from the vacuum sector. However, in terms of the
$N=2$ algebra, there are infinitely many different Ishibashi
states. In fact, because of \eqref{eq:3} we have an $N=2$ Ishibashi
state for each $(n,Q)$, where $(n,Q)=(0,0)$ or $n\in\mN$ and 
$Q=\pm 1$. In order to introduce a uniform notation, we denote these
$N=2$ Ishibashi states as  
\begin{equation}
  \label{eq:14}
  \Iket{n,\eta}=\begin{cases} \Iket{(n,1),\eta} & \text{for } n>0,\\
    \Iket{(0,0),\eta} & \text{for } n=0, \\
    \Iket{(|n|,-1),\eta} & \text{for } n<0.
    \end{cases}
\end{equation}
We fix the normalisation of these Ishibashi states by demanding that 
the usual Neumann U(1) boundary state 
\begin{equation}
  \label{eq:15}
    \Bket{N,\eta}^{\text{free}}=
    \mathcal{N}
    \exp{\left[-\sum_{n>0}\frac{1}{n}\left(\alpha_{-n}\tilde{\alphabar}_{-n}
          + \alphabar_{-n}\tilde{\alpha}_{-n} \right)
        - i \eta \sum_{r>0}\left( \psi^+_{-r}\tilde{ \psi}^-_{-r} 
          + \psi^+_{-r}\tilde{ \psi}^-_{-r}\right)\right]}\ket{0} \,,
\end{equation}
where $\mathcal{N}$ is a normalisation constant, equals
\begin{equation}
  \label{eq:13}
  \Bket{N,\eta}^{\text{free}}= \mathcal{N}\, 
  \sum_{n\in \mZ} \Iket{n,\eta}\,.
\end{equation}
Clearly, this is not the most general $N=2$ boundary state since the
coefficients in front of the different $N=2$ Ishibashi states do not
all have to be equal. The most general ansatz for a $N=2$ boundary
state from the NS-NS sector is 
\begin{equation}
  \label{eq:13a}
  \Bket{C,\eta}^{N=2}= 
  \sum_{n\in \mZ} C_n\, \Iket{n,\eta}\,,
\end{equation}
where $C_n$ are constants that need to be determined, using the usual
consistency conditions, in particular Cardy's condition
\cite{Cardy:1989ir}. 

Given that the $N=2$ Ishibashi states are labelled by the integers, it
is very suggestive that (up to an overall normalisation), the $C_n$
should simply equal 
\begin{equation}
\label{eq:ansatz}
C_n = e^{in\phi} \,,
\end{equation}
where $\phi$ is an angular variable. Thus our ansatz for the boundary
state is 
\begin{equation}
  \label{eq:18}
  \Bket{N,\phi,\eta} = \mathcal{N}_{NS}(\phi)
\sum_{n\in\mZ} e^{in\phi}\, \Iket{n,\eta}\,,
\end{equation}
where $\mathcal{N}_{NS}(\phi)$ is a normalisation constant that will be
determined below. It is clear that if this family of boundary states
is in fact consistent for all $\phi$, then it is already the most
general class of boundary states. This is simply a consequence of the
fact that we can invert (essentially by Fourier transformation) the
above relation to write every Ishibashi state $\Iket{n,\eta}$ as an
integral over the boundary states $\Bket{N,\phi,\eta}$. Any 
boundary state of the form \eqref{eq:13a} can therefore be
expressed as a linear combination (or possibly an integral) of these
boundary states. In this sense the above boundary states are therefore
generating all boundary states (if they are indeed consistent).

On the other hand, given the structure of the singular vectors
\eqref{eq:17}, comparison of \eqref{eq:13} with \eqref{eq:15} suggests
that this ansatz simply corresponds to switching on an electric field
on the usual Neumann brane; in terms of the gluing conditions of the
free fields this means that  
\begin{subequations}
  \label{eq:92}
  \begin{align}
  (\alpha_n + e^{-i\phi}\tilde{\alpha}_{-n}) 
  \Bket{N,\phi,\eta}^{\text{free}} &=0\,, \\
  (\alphabar_n + e^{i\phi}\tilde{\alphabar}_{-n})
   \Bket{N,\phi,\eta}^{\text{free}} &=0\,, \\
  (\psi^+_r + i\eta e^{-i\phi}
\tilde{\psi}^+_{-r})\Bket{N,\phi,\eta}^{\text{free}} &=0\,, \\
  (\psi^-_r + i\eta e^{i\phi}
\tilde{\psi}^-_{-r})\Bket{N,\phi,\eta}^{\text{free}} &=0\,. 
\end{align}
\end{subequations}
It is clear from \eqref{eq:89} that this automorphism acts trivially
on the $N=2$ generators, and therefore that it leaves the $N=2$ gluing 
conditions unmodified. 

In the following section we shall confirm that this expectation is indeed
correct; more specifically, we shall show that the cylinder amplitudes
of \eqref{eq:18} agree with those of a Neumann brane with an electric
field. This will then also allow us to determine the correct
normalisation constants $\mathcal{N}_{NS}(\phi)$. 

\section{Calculation of the amplitude}
\label{sec:amplitude}

In this section we want to calculate the overlap of two NS-NS boundary
states of the form \eqref{eq:18}. Recall that our normalisation
convention for the $N=2$ Ishibashi states implies that 
\begin{equation}
  \label{eq:19}
  \Ibra{n}q^{L_0 + \tilde{L}_0 -
  \frac{1}{4}}\Iket{n'} = \chi_n(q)\, \delta_{n,n'} \,,
\end{equation}
where $\chi_n$ denotes the appropriate $N=2$ character. We shall first
consider the situation where the two spin structures are the same (but
the phases $\phi$ may differ)
\begin{subequations}
  \label{eq:20}
\begin{align}
  \mathcal{A}^+_{\phi_1,\phi_2} &=\Bbra{N,\phi_1,\eta}q^{L_0 +
    \tilde{L}_0 - \frac{1}{4}}\Bket{N,\phi_2,\eta} \\
  & = \mathcal{N}_{NS}(\phi_1) \, \mathcal{N}_{NS}(\phi_1)\, 
\sum_{n\in \mZ}e^{in(\phi_1-\phi_2)} \chi_n(q)\,.
\end{align}
\end{subequations}
If we set $\Delta:=\phi_1-\phi_2$ we can rewrite the sum in
\eqref{eq:20} as
\begin{multline}
    \label{eq:21}
  \sum_{n\in\mZ}e^{i n\Delta} \chi_n(q) = \chi_0(q) +
  \sum_{n>0}(e^{in\Delta} + e^{-in\Delta}) \chi_n(q) \\
  = q^{-\frac{1}{4}}\prod_{n=1}^{\infty}\frac{(1+q^{2n-1})^2}{(1-q^{2n})^2}
  \left[1+2\sum_{n=1}^\infty [\cos(n\Delta) -
    \cos((n-1)\Delta)]\frac{q^{2n-1}}{1+q^{2n-1}}
  \right] \,.
\end{multline}
As was explained in the previous section, we expect this amplitude to
agree with that of D-branes with electric flux (labelled by $\phi_1$
and $\phi_2$, respectively). The cylinder amplitude for two such
branes is given by 
(see for example \cite{Bachas:1998rg,Bachas:1992bh})
\begin{equation}
  \label{eq:77}
  \mathcal{A}^{\text{electric}}=
    \left(\cos{\tfrac{\phi_1}{2}}\cos{\tfrac{\phi_2}{2}}\right)^{-1}    
    q^{-\frac{1}{4}}
    \frac{\prod_{n=1}^{\infty}    
    (1+2\cos(\Delta)q^{2n-1} + 
    q^{4n-2})}{\prod_{n=1}^{\infty}(1-2\cos(\Delta)q^{2n} + q^{4n})}\,.
\end{equation}
In terms of \vth-functions formula \eqref{eq:77} can be written as
\begin{equation}
  \label{eq:83}
    \mathcal{A}^{\text{electric}}=
  \frac{2\sin{\left(\tfrac{\Delta}{2}\right)}}
  {\cos{\tfrac{\phi_1}{2}}\cos{\tfrac{\phi_2}{2}}} 
  \frac{\vt{3}{\frac{\Delta}{2},q}}{\vt{1}{\frac{\Delta}{2},q}}\,,
\end{equation}
which makes its modular properties manifest. (The relevant 
\vth-functions are defined in appendix \ref{sec:theta}, where their
modular properties are also reviewed.) It is a non-trivial fact that
these two expressions, {\it i.e.} \eqref{eq:21} and \eqref{eq:77}, 
are actually equal (up to normalisation); this is proven in appendix
\ref{sec:proof}. With the help of this result, and keeping track of
the various normalisation constants, we then find that 
\begin{equation}
  \label{eq:30}
\mathcal{A}^+_{\phi_1,\phi_2}=
 2\mathcal{N}_{NS}(\phi_1)\mathcal{N}_{NS}(\phi_2)
\sin{\left(\tfrac{\Delta}{2}\right)} 
 \frac{\vt{3}{\frac{\Delta}{2},q}}{\vt{1}{\frac{\Delta}{2},q}} \,,
\end{equation}
where the $\mathcal{N}_{NS}(\phi_i)$ are the same constants that
appeared in \eqref{eq:18}.

\subsection{GSO projection}
\label{sec:gso-projection}

Up to now we have only considered one spin structure, but in order to
obtain spacetime supersymmetric branes, we need to impose a
GSO-projection, and this will require that we consider both spin
structures together. The GSO-invariant NS-NS boundary states can be
constructed by the following standard method (see for example
\cite{Gaberdiel:2000jr}). Strictly speaking, when considering the
GSO-projection we have to specify the superconformal field theory that
makes up the remaining central charge and apply the GSO-projection to
the full theory. In order to avoid having to make a choice for this
additional theory we restrict ourselves in the following to
GSO-projecting the boundary states of our $c=3$ theory separately; the
modifications that arise when including the remaining degrees of
freedom in the full theory are straightforward. Let us assume that the
relevant projection for the $c=3$ theory in the NS-NS sector is simply  
\begin{equation}
  \label{eq:38}
  P_{GSO} = \tfrac{1}{4}(1 + (-1)^F)(1 + (-1)^{\tilde{F}})\,.
\end{equation}

By definition, the action of the world-sheet fermion number operator
on the NS-NS ground state is given by 
\begin{equation}
  \label{eq:35}
  (-1)^{\tilde{F}}\ket{0,0} = (-1)^F\ket{0,0} = -\ket{0,0}\,. 
\end{equation}
$(-1)^F$ anticommutes with fermionic modes, and its action on the  
singular vectors given by \eqref{eq:17} is therefore
$$
(-1)^F v_{(n)}^{(\pm)} = +v_{(n)}^{(\pm)}\,.
$$
The GSO-invariant combinations of Ishibashi states are therefore
\begin{equation}
  \label{eq:37}
  \Iket{n}=
  \begin{cases}
    \tfrac{1}{\sqrt{2}}(\Iket{n,\eta} + \Iket{n,-\eta}), \qquad
    \text{for }n\neq0 \\
    \tfrac{1}{\sqrt{2}}(\Iket{0,\eta} - \Iket{0,-\eta}), \qquad\,
    \text{for }n=0\,. \\
  \end{cases}
\end{equation}
Our ansatz for the total boundary state is thus
\begin{equation}
  \label{eq:18n}
  \Bket{N,\phi} = \mathcal{N}_{NS}(\phi)
\sum_{n\in\mZ} e^{in\phi}\, \Iket{n}\,,
\end{equation}
where $\mathcal{N}_{NS}(\phi)$ is a normalisation constant that will
be determined below.

The cylinder diagram of two such GSO-invariant boundary states is
now
\begin{equation}
\Bbra{N,\phi_1}q^{(L_0 + \tilde{L}_0 -
    \frac{1}{4})}\Bket{N,\phi_2} = 
\left( \mathcal{A}^+_{\phi_1,\phi_2} + 
\mathcal{A}^-_{\phi_1,\phi_2} \right) \,,
\end{equation}
where $\mathcal{A}^+_{\phi_1,\phi_2}$ is given in 
\eqref{eq:30}, while $\mathcal{A}^-_{\phi_1,\phi_2}$ is the
contribution from $\eta_1=-\eta_2$. Because of the sign structure of 
\eqref{eq:37} this equals
\begin{multline*}
 \mathcal{A}^-_{\phi_1,\phi_2} =
 \mathcal{N}_{NS}(\phi_1)\mathcal{N}_{NS}(\phi_2) \times 
 \\
\times 
 q^{-1/4}\prod_{n=1}^\infty\frac{(1-q^{2n-1})^2}{(1-q^{2n})^2}\left[-1 -
    \frac{2q}{1-q} + 2\sum_{n}\cos(n\Delta)\left(\frac{q^{2n-1}}{1 -
    q^{2n-1}} - \frac{q^{2n+1}}{1 -
    q^{2n+1}}\right)\right]\,.
\end{multline*}
This can be expressed, in terms of $\vartheta$-functions, as
\begin{equation}
  \label{eq:93}
\mathcal{A}^-_{\phi_1,\phi_2} =
    -2\mathcal{N}_{NS}(\phi_1)\mathcal{N}_{NS}(\phi_2)\sin{ \left(
    \tfrac{\Delta}{2}\right)} 
    \frac{\vt{4}{\frac{\Delta}{2},q}}{\vt{1}{\frac{\Delta}{2},q}}\,. 
\end{equation}
This follows by a similar identity as the one used before, and can be
proven by the method of appendix \ref{sec:proof}.

\section{The Ramond Sector}
\label{sec:ramond-sector}

Next we want to perform the analogous analysis in the R-R
sector. Most of the analysis will be the same as for the NS-NS sector.

First of all, if ${\bf p}\ne 0$, there is precisely one $N=2$
Ishibashi state (for each choice of $\eta$), which therefore agrees
with the usual Dirichlet Ishibashi state. Again these Ishibashi states
appear in the boundary states for the usual Dirichlet branes, and do
not give rise to any interesting new D-branes.

As in the NS-NS sector, the situation is more interesting in the
vacuum sector. By the usual argument we get one Ishibashi state for
each irreducible representation that appears in \eqref{eq:3R}. It is
convenient to label the Ishibashi states by $\Iket{n,\pm,\eta}$ for
$n>0$, as well as $\Iket{0,+,\eta}$ and $\Iket{0,-,\eta}$. (Thus for
example the Ishibashi state $\Iket{n,\pm,\eta}$ has the expansion
\begin{subequations}
\begin{eqnarray}
\Iket{n,+,\eta} & \sim &  
w^{(+,+)}_{(n)} \otimes \tilde{w}^{(-,-)}_{(n)} + \cdots \label{Ish1}
\\
\Iket{n,-,\eta} & \sim &  
w^{(-,-)}_{(n)} \otimes \tilde{w}^{(+,+)}_{(n)} + \cdots\,.) \label{Ish2}
\end{eqnarray}  
\end{subequations}
Furthermore, as in the NS-NS sector, we fix its normalisation by
demanding that the usual Neumann boundary state is simply given as 
\begin{equation}
\Bket{N,\phi,\eta}^{\text{free}} = {\cal N} \, 
\left[ \Iket{0,+,\eta} + \Iket{0,-,\eta} 
+ \sum_{n\in\mN} \left(\Iket{n,+,\eta} + \Iket{n,-,\eta} \right)
\right]\,,
\end{equation}
where ${\cal N}$ is some normalisation constant. As before in the
NS-NS sector, this is not the most general $N=2$ boundary state; in
order to obtain a spacetime supersymmetric brane, we need to mimic the
construction in the NS-NS sector, and we therefore expect that we
should simply switch on an electric field on the world-volume of the
brane. Given the structure of the $N=2$ singular vectors in terms of
the free fields, this suggests the ansatz
\begin{eqnarray}
\label{eq:41}
\Bket{R,\phi,\eta} & = & \mathcal{N}_R(\phi) \biggl(
e^{\frac{i}{2}\phi}\Iket{0,+,\eta} +
e^{-\frac{i}{2}\phi}\Iket{0,-,\eta} +  \\ \nonumber 
& & \qquad 
\sum_{n\in \mN} e^{i(n+\frac{1}{2})\phi} \Iket{n,+,\eta} +
e^{-i(n+\frac{1}{2})\phi} \Iket{n,-,\eta}\bigg)\,.
\end{eqnarray}
By the same reasoning as before in the NS-NS sector, these boundary
states will account for all Ishibashi states (if they are indeed
consistent) since we can invert this relation and write the Ishibashi
states in terms of these boundary states. 

The overlap between two such boundary states now becomes
\begin{subequations}
  \label{eq:51}
\begin{align}
  \mathcal{A}^R_{\phi_1,\phi_2} &= \Bbra{R,\phi_1,\eta}q^{L_0 +
    \tilde{L}_0 + \frac{1}{8}}\Bket{R,\phi_2,\eta} \\
& = \mathcal{N}_R(\phi_1) \mathcal{N}_R(\phi_2)\,
\left( 2 \cos (\Delta /2) \, \chi^{\text{R}}_0(q) 
+ 2\sum_{n=1}^\infty\cos((n+1/2)\Delta)\chi_n^{\text{R}}(q)\right)\,, 
\end{align}
\end{subequations}
where $\Delta = \phi_1 - \phi_2$ as before, and $\chi_n^{\text{R}}(q)$
denotes the character \eqref{eq:7}. 
Again, we can express the amplitude in terms of \vth-functions (using
a relation similar to the one proven in appendix \ref{sec:proof}), 
and we thus obtain
\begin{equation}
  \label{eq:54}
  \mathcal{A}^R_{\phi_1,\phi_2} =
  2 \mathcal{N}_R(\phi_1) \, \mathcal{N}_R(\phi_2)\,
\sin{\left(\tfrac{\Delta}{2}\right)} \,
  \frac{\vt{2}{\frac{\Delta}{2},q}}
  {\vt{1}{\frac{\Delta}{2},q}} \,.
\end{equation}

\subsection{GSO projection}
\label{sec:r-gso-projection}

As before, we need to make sure that the boundary state is invariant
under the GSO-projection. In the presence of fermionic zero modes,
this requires (as always) a little bit of care. We choose the
convention that 
\begin{equation}\label{fermnum}
(-1)^F \ket{\tfrac{1}{8},\tfrac{1}{2}} = +
\ket{\tfrac{1}{8},\tfrac{1}{2}} \,, \qquad
(-1)^{\tilde{F}} \ket{\tfrac{1}{8},\tfrac{1}{2}} = +
\ket{\tfrac{1}{8},\tfrac{1}{2}} \,.
\end{equation}
It is always possible to choose the normalisation of the Ishibashi
states so that 
\begin{equation}
(-1)^F\Iket{n,Q,\eta}= \Iket{n,Q,-\eta} \,,
\end{equation}
but then the action of $(-1)^{\tilde{F}}$ is unambiguous. Given the
form of the Ishibashi states \eqref{Ish1}, \eqref{Ish2}, as well as
our conventions about the action of the fermion number operators
\eqref{fermnum} it is then easy to see that 
\begin{equation}
(-1)^{\tilde{F}}\Iket{n,Q,\eta}= -\Iket{n,Q,-\eta} \,.
\end{equation}
(The same also holds for the two Ishibashi states that come 
from the R ground states.) In order to obtain consistent boundary
states we have to chose the GSO projector in the R sector as a type
IIA GSO projector\footnote{This does not mean that these boundary
states only exist in the complete ten-dimensional theory if it is type
IIA; however it does give constraints on whether this Neumann boundary
condition for the $c=3$ sector is consistent within the context of
the complete theory. The fact that a IIA projection appears simply 
reflects the fact that we are describing a D2-brane here.}
\begin{equation}
  \label{eq:58}
   P_{GSO}=\tfrac{1}{4}(1 + (-1)^F)(1 - (-1)^{\tilde{F}})
\end{equation}
and the GSO invariant states are then given by
\begin{equation}
  \label{eq:59}
  \Iket{n,\pm}_{RR} =
  \tfrac{1}{\sqrt{2}}(\Iket{n,\pm,+} -   \Iket{n,\pm,-})\,,
\end{equation}
with a similar formula for $\Iket{0,\pm}$. The GSO-invariant ansatz 
for our boundary state is then simply
\begin{equation}
  \Bket{R,\phi} = \mathcal{N}_R(\phi) \left(
e^{\frac{i}{2}\phi}\Iket{0,+} +
  e^{-\frac{i}{2}\phi}\Iket{0,-} +
  \sum_{n\in \mN} e^{i(n+\frac{1}{2})\phi}\Iket{n,+} +
  e^{-i(n+\frac{1}{2})\phi}\Iket{n,-}\right)\,.
\end{equation}
The calculation of the cylinder amplitude involving these
GSO-invariant boundary states is trivial since the overlap between two
boundary states with opposite $\eta$ vanishes. Thus the formula simply
reproduces again \eqref{eq:54}.

\section{The open string amplitude and normalisation}
\label{sec:normalisation}

Finally, in order to fix the normalisation constants and to check that
our results are consistent with the open string results, we perform a 
modular transformation on the amplitudes to the open string sector. We
make the ansatz for our boundary state to be simply 
\begin{equation}
\Bket{\phi} = \frac{1}{\sqrt{2}}
\Bigl(\Bket{N,\phi} + i \epsilon \Bket{R,\phi}\Bigr) \,,
\end{equation}
where $\epsilon=\pm 1$ distinguishes a brane from an anti-brane. Using
the results we have derived in the previous sections, it is easy to
see that the cylinder amplitude between two such branes gives rise, in
the open string description, to the amplitude
\begin{subequations}
\begin{align}
  \label{ergeb}
  {\cal A} & = i {\cal N}_{NS}(\phi_1) {\cal N}_{NS}(\phi_2) 
\sin{ \left(\tfrac{\Delta}{2}\right)} 
\left[ 
\frac{\vt{3}{\frac{\Delta\ttau}{2},\qt}}
{\vt{1}{\frac{\Delta\ttau}{2},\qt}}
- \frac{\vt{2}{\frac{\Delta\ttau}{2},\qt}}
{\vt{1}{\frac{\Delta\ttau}{2},\qt}}\right]
\\
&  \qquad - i  {\cal N}_{R}(\phi_1) {\cal N}_{R}(\phi_2) 
\sin{\left(\tfrac{\Delta}{2}\right)} 
\frac{\vt{4}{\frac{\Delta\ttau}{2},\qt}}
{\vt{1}{\frac{\Delta\ttau}{2},\qt}} \,,
\end{align}
\end{subequations}
where $\ttau=-\frac{1}{\tau}$ and $\qt=e^{i\pi\ttau}$. In deriving
this formula we have used the modular transformation properties of
the \vth-functions that are detailed in the appendix.  

The above formula now suggests that we should set 
$\mathcal{N}_{NS}(\phi)=\mathcal{N}_{R}(\phi)$; furthermore, since
these amplitudes are the same as for an open string in the presence of
a background electric field, the analysis of 
\cite{Abouelsaood:1987gd} (see also 
\cite{Bachas:1998rg,DiVecchia:1999rh,DiVecchia:1999fx}) suggests that 
\begin{equation}
\label{eq:33}
\mathcal{N}_{NS}(\phi) = \mathcal{N}_{R}(\phi) = 
\frac{1}{\cos{\frac{\phi}{2}}}\,. 
\end{equation}
For example, the first term in (\ref{ergeb}) then becomes after a
short calculation
\begin{equation}
\label{eq:34}
\frac{1}{2}\left(\tan{\left(\tfrac{\phi_1}{2}\right)}
    - \tan{\left(\tfrac{\phi_2}{2}\right)} \right)
  \frac{\qt^{\frac{\Delta}{2\pi}}}{1-\qt^{\frac{\Delta}{\pi}}}
  \qt^{-1/4} \prod_{n=1}^\infty
  \frac{(1+e^{i\Delta\ttau}\qt^{2n-1})(1+e^{-i\Delta\ttau}\qt^{2n-1})} 
  {(1-e^{i\Delta\ttau}\qt^{2n})(1-e^{-i\Delta\ttau}\qt^{2n})}\,.
\end{equation}
This is the correct mode expansion for (one half of) the open string
NS sector \cite{Bachas:1998rg,DiVecchia:1999rh,DiVecchia:1999fx}. The
other terms work similarly; in particular, the contribution from the
RR boundary states gives rise to one half of the open string NS sector
with the insertion of $(-1)^F$, and thus combines with \eqref{eq:34}
to give the correct GSO-projection.

\section{Conclusion}
\label{sec:conclusion}

In this paper we have constructed the most general $N=2$
superconformal boundary states for the theory of two free
(uncompactified) bosons and fermions. We have shown that the only
$N=2$ D-branes for this theory are the usual Dirichlet
branes, as well as the Neumann branes with an electric field. 
In the process of identifying the $N=2$ boundary states with Neumann
branes with electric flux we were led to prove some non-trivial
\vth-function identities that may be interesting in their own right.

Our analysis implies in particular that for this theory the
usual Neumann and Dirichlet branes already account for {\it all}
$N=2$ D-branes. This is in marked contrast to the results in the
non-supersymmetric case, or the case with $N=1$ supersymmetry, where
there exist conformal or $N=1$ superconformal D-branes that do not
have an interpretation in terms of Neumann or Dirichlet branes
\cite{Gaberdiel:2001zq}. 

On the other hand, in contradistinction to the cases mentioned above,
the boundary states we have constructed in this paper preserve
$N=2$ worldsheet supersymmetry, and the corresponding D-branes are
therefore spacetime supersymmetric and stable. Our results therefore
imply that the usual Neumann and Dirichlet branes already account for
all spacetime supersymmetric D-branes in this simple example. This is
in some sense reassuring since until now mostly these D-branes have
been considered.

There are indications that there exist other theories (for example
certain orbifold theories \cite{CG}) for which the 
supersymmetric D-branes charges are not just generated by the usual
Neumann and Dirichlet branes. One should hope that the techniques of
this paper will help to construct those branes.

\section*{Acknowledgements}
\label{sec:acknoledgements}

We would like to thank Terry Gannon, Kevin Graham, Peter Kaste, Tako
Mattik, Andreas  
Recknagel and Daniel Roggenkamp for interesting discussions. This 
research is partially supported by the Swiss National Science
Foundation. The work of H.K. is also supported by a scholarship of the
Marianne und Dr. Fritz Walter Fischer-Stiftung and a
Promotionsstipendium of the Deutscher Akademischer Austauschdienst
(DAAD). 

\appendix

\section{Singular vectors}
\label{sec:singular-vectors}

In this appendix we want to show that the  vectors given in
equation \eqref{eq:17}, namely
\begin{subequations}
  \label{eq:79}
\begin{align}
  v^{(+)}_{(n)} &= (\alpha_{-1})^n\psi^{+}_{-1/2}\ket{0} \\
  v^{(-)}_{(n)} &= (\alphabar_{-1})^n\psi^{-}_{-1/2}\ket{0} 
\end{align}
\end{subequations}
are actually singular vectors for the $N=2$ algebra.

The mode expansions of the $N=2$ operators in terms of the free fields
are given in equations \eqref{eq:89}. We need to prove that 
$S_m \, v =0$ for $m>0$, where $S$ stands for any generator of the
$N=2$ algebra. Consider for example the situation where $S=L$, 
$$
L_m\, v^{(+)}_{(n)}, \quad m>0
$$ 
which, in terms of the free field modes, is given by
\begin{equation}
  \label{eq:95}
  \left(\sum_{n} :\alpha_{m-n}\alphabar_{n}: +
  \frac{1}{2}\sum_{s}(2s-m):\psi^{-}_{m-s}\psi^+_s:\right) 
     (\alpha_{-1})^n\psi^{+}_{-1/2}\ket{0}\,.
\end{equation}
Let us focus on the bosonic part first. Due to the (anti-)commutation
relations given in \eqref{eq:87}, it is clear that
all products in the expansion of $L_m$ which do not contain
$\alphabar_{1}$ contain an operator with positive mode number which
commutes with the creation operators that appear in the singular
vector and annihilates the ground state.  Therefore the only terms we
have to consider more closely are of the form
$\alpha_{m-1}\alphabar_1$. By assumption $m\geq 1$ and therefore
$m-1\geq 0$. Thus any term of this form contains an annihilation
operator which again commutes with the operators in the singular
vector, and thus annihilates the  ground state. This shows that the
bosonic part of $L_m$ annihilates $v_n$. The argument for the
fermionic part runs along the same lines. The same argument works
also for the other $N=2$ generators, except that for the supercharges
one has to use additionally that  
$(\psi_{-1/2}^+)^2=0$.

\section{Various $\vartheta$ function definitions and identities}
\label{sec:theta}

This is a list of $\vartheta$ function definitions and identities,
collected at various points; in particular \cite{bateman} and
\cite{whittaker58} are useful references. 

\paragraph{Definition of \vth-functions}
\label{sec:definition-theta}

The \vth-functions are defined as 
\begin{subequations}
  \label{eq:62}
\begin{align}
  \vt{1}{z,q}&=2\sum_{n=0}^{\infty}(-1)^nq^{(n+\frac{1}{2})^2}\sin(2n+1)z\,,
  \\
  \vt{2}{z,q}&=2\sum_{n=0}^{\infty}q^{(n+\frac{1}{2})^2}\cos(2n+1)z \,,\\
  \vt{3}{z,q}&=1 + 2\sum_{n=1}^{\infty}q^{n^2}\cos2nz \,,\\
  \vt{4}{z,q}&=1 + 2\sum_{n=1}^{\infty}(-1)^nq^{n^2}\cos2nz \,,
\end{align}
\end{subequations}
where $q=e^{i \pi \tau}$.
They are related among each
other by 
\begin{subequations}
  \label{eq:64}
\begin{align}
  \vt{1}{z,q}&=-ie^{iz+\frac{1}{4}i\pi\tau}\vt{4}{z+\tfrac{1}{2}\pi\tau,q}\,,
  \\
  \vt{2}{z,q}&=\vt{1}{z+\tfrac{1}{2}\pi,q} \,,\\
  \vt{3}{z,q}&=\vt{4}{z+\tfrac{1}{2}\pi,q}\,.
\end{align}
\end{subequations}
The \vth-functions  can be
written as infinite products as follows 
\begin{subequations}
  \label{eq:63}
\begin{align}
  \vt{1}{z,q}&=2q^{\frac{1}{4}}\sin z\,
  G\prod_{n=1}^{\infty}(1-2q^{2n}\cos{2z} + q^{4n}) \,,\\
  \vt{2}{z,q}&=2q^{\frac{1}{4}}\cos z\,
  G\prod_{n=1}^{\infty}(1+2q^{2n}\cos{2z} + q^{4n}) \,,\\
  \vt{3}{z,q}&= G\prod_{n=1}^{\infty}(1+2q^{2n-1}\cos{2z} + q^{4n-2}) \,,\\
  \vt{4}{z,q}&= G\prod_{n=1}^{\infty}(1-2q^{2n-1}\cos{2z} + q^{4n-2}) \,,
\end{align}
\end{subequations}
where $G$ is given by 
$$
G=\prod_{n=1}^{\infty}(1-q^{2n})\,.
$$
We further define $\phi(z,q)$ by $\vt{1}{z,q}=\sin z\, \phi(z,q)$. 

\paragraph{Periodicities of \vth-functions}
\label{sec:periodicities}

The \vth-functions are quasi doubly-periodic in $z$ with period $\pi$ and
$\pi\tau$. Under transformation of $z$ by a period they pick up the
factors
\begin{center}
  \begin{tabular}[c]{|c|c|c|c|c|}
\hline
&$\vt{1}{z,q}$&$\vt{2}{z,q}$&$\vt{3}{z,q}$&$\vt{4}{z,q}$ \\
\hline
$\pi$ & $-1$ & $-1$ & 1 & 1 \\
\hline
$\pi\tau$& $-N$ & $N$ & $N$ & $-N$\\
\hline
\end{tabular}
\end{center}
where $N=q^{-1}e^{-2iz}$.

By considering the definitions of the \vth-functions we can see at
once that $\vt{1}{z,q}$ is an odd function in $z$ whereas all other
\vth-functions are even in $z$.

\paragraph{The zeros of \vth-functions}
\label{sec:zeros}

If $\vt{i}{z,q}$ has a zero at $z_0$ the quasi-periodicity implies it
has also zeros at
$$
z_0 + m\pi + n\pi\tau\,, \qquad  n,m\in \mZ\,.
$$
It can be shown that in the fundamental domain with corners
$\{0,\pi,\pi\tau,\pi+\pi\tau\}$ the zeros are given by

\begin{center}
\begin{tabular}{|c|c|c|c|}
\hline
$\vt{1}{z,q}$&$\vt{2}{z,q}$&$\vt{3}{z,q}$&$\vt{4}{z,q}$ \\
\hline
$0$ & $\frac{1}{2}\pi$ & $\frac{1}{2}\pi + \frac{1}{2}\pi\tau$ & $
\frac{1}{2}\pi\tau$\\
\hline 
\end{tabular}
\end{center}

\paragraph{\vth-function identities}
\label{sec:vth-funct-ident}

\vth-functions are connected via a vast amount of identities of which
we will only list very few here. 
Two remarkable identities are 
\begin{equation}
  \label{eq:65}
  \vartheta_{1}'(0,q) = \vt{2}{0,q}\vt{3}{0,q}\vt{4}{0,q} \,,
\end{equation}
where the prime denotes a differentiation with respect to $z$ and the
well known identity
\begin{equation}
  \label{eq:66}
  \vt{2}{0,q}^4 + \vt{4}{0,q}^4 = \vt{3}{0,q}^4 \,,
\end{equation}
which can be generalised to
\begin{equation}
  \label{eq:67}
  \vt{2}{z,q}^4 + \vt{4}{z,q}^4 = \vt{1}{z,q}^4 +\vt{3}{z,q}^4\,.
\end{equation}
From formula \eqref{eq:65} we can deduce the interesting product
identity 
\begin{equation}
  \label{eq:68}
  1 = \prod_{n=1}^{\infty}(1+q^{2n})\prod_{n=1}^{\infty}(1+q^{2n-1})
      \prod_{n=1}^{\infty}(1-q^{2n-1}) 
\end{equation}
by writing out the \vth-functions as products.

\paragraph{Modular transformations}
\label{sec:modul-transf}

It is sometimes convenient to write the \vth-functions as functions of
$\tau$ and $z$, rather than as functions of $q$ and $z$. In order to
distinguish this from the previous definition of the \vth-functions we
use the symbol $\vt{i}{z|\tau}$ in that case. In particular, this is
convenient when discussing the modular transformation properties of
the \vth-functions:

\begin{center}
\begin{tabular}[c]{|c|c|c|c|c|}
\hline
\rule[-2mm]{0mm}{6mm}& $\vt{1}{z|\tau}$ &
$\vt{2}{z|\tau}$&$\vt{3}{z|\tau}$&$\vt{4}{z|\tau}$\\
\hline
\rule[-1.5mm]{0mm}{6mm}$\tau'=\tau+1$& 
$ e^{-\frac{\pi i}{4}} \vt{1}{z|\tau'}$ & $ 
e^{-\frac{\pi i}{4}} \vt{2}{z|\tau'} $&$ 
\vt{4}{z|\tau'} $&$  \vt{3}{z|\tau'}$\\
\hline
\rule[-2mm]{0mm}{6mm}$\tau'= \tfrac{-1}{\tau}$& 
$-iA \vt{1}{z\tau'|\tau'} $&
$ A\vt{4}{z\tau'|\tau'} $ &
$ A\vt{3}{z\tau'|\tau'} $ &
$ A\vt{2}{z\tau'|\tau'} $ \\
\hline
\end{tabular}
\end{center}
where $A=(-i\tau)^{-1/2} e^{\frac{iz^2\tau'}{\pi}}$.

\section{Proof of the \vth-function identity}
\label{sec:proof}

We want to prove that 
\begin{equation}
  \label{eq:23}
  \mathcal{A}=q^{-\frac{1}{4}}\frac{\prod_{n=1}^{\infty}
    (1+2\cos(\Delta)q^{2n-1} + 
    q^{4n-2})}{\prod_{n=1}^{\infty}(1-2\cos(\Delta)q^{2n} + q^{4n})}
\end{equation}
is equal to 
\begin{equation}
\label{eq:86}
\mathcal{A}_{\phi_1,\phi_2}
=q^{-\frac{1}{4}}\prod_{n=1}^{\infty}\frac{(1-q^{2n-1})^2}{(1-q^{2n})^2}
\left[1+2\sum_{n=1}^\infty [\cos(n\Delta) -
\cos((n-1)\Delta)]\frac{q^{2n-1}}{1+q^{2n-1}}
\right] \,.
\end{equation}
The former expression can be easily written in terms of 
$\vartheta$-functions,
\begin{equation}
  \label{eq:22}
  \mathcal{A}=
  2\sin{\left(\tfrac{\Delta}{2}\right)}
  \frac{\vt{3}{\frac{\Delta}{2},q}}{\vt{1}{\frac{\Delta}{2},q}} \,.
\end{equation}
Similarly, we can easily write the prefactor of \eqref{eq:86}
in this manner
\begin{equation}
  \label{eq:25}
  \mathcal{A}_{\phi_1,\phi_2} =
  2\frac{\vt{3}{0,q}}{\phi(0,q)}
  \left[1+2\sum_{n=1}^\infty [\cos(n\Delta) - \cos((n-1)\Delta)]
    \frac{q^{2n-1}}{1+q^{2n-1}} 
  \right] \,,
\end{equation}
where $\vt{1}{z,q} = \sin{(z)}\phi(z,q)$. To show the equality of the
two expressions we will treat them as analytic functions in the first
variable and show that their periodicities, zeros and poles are
identical. So, what we want to prove is 
\begin{eqnarray}
  \label{eq:26}
  \frac{\vt{3}{z,q}}{\vt{1}{z,q}}& = & \frac{1}{\sin(z)} 
  \frac{\vt{3}{0,q}}{\phi(0,q)}
  \left[1+2\sum_{n=1}^\infty [\cos(2nz) - \cos(2(n-1)z)]
    \frac{q^{2n-1}}{1+q^{2n-1}}
  \right] \nonumber \\
& = & \frac{1}{\sin(z)} \frac{\vt{3}{0,q}}{\phi(0,q)} \Sigma 
\end{eqnarray}
where we have divided both sides by $2\sin(z)$, and where $\Sigma$ is
the expression in brackets
\begin{equation}
\Sigma = 1+2\sum_{n=1}^\infty [\cos(2nz) - \cos(2(n-1)z)]
    \frac{q^{2n-1}}{1+q^{2n-1}} \,.
\end{equation}
The zeros of the left hand side of \eqref{eq:26} are given by 
$\tfrac{\pi}{2} + \tfrac{\pi\tau}{2}$. If we insert this value for $z$
into the expression for $\Sigma$ we get 
\begin{equation}
  \label{eq:27}
  1 + \sum_{n=1}^{\infty}(-1)^n[q^n + q^{n-1}]=0 \,,
\end{equation}
which shows that $\tfrac{\pi}{2} + \tfrac{\pi\tau}{2}$ is indeed a
zero of the right hand side. In order to show that the right hand side
of equation \eqref{eq:26} does not have any additional zeros we use
the trigonometric identity 
\begin{equation}
\cos(2nz) - \cos(2(n-1)z) = - 2 \sin(z) \, \sin((2n-1)z)
\end{equation}
to rewrite $\Sigma$ as 
\begin{eqnarray}
\Sigma & = & 2 \sin(z) \,
\left[ \frac{1}{2 \sin(z)} - 2 \sum_{n=1}^{\infty} 
\sin(2n-1)z \, \frac{q^{2n-1}}{1 + q^{2n-1}} \right] \label{C7} \\
& = &  2 \sin(z) \,
\left[- 2\, i \sum_{n=1}^{\infty} 
\frac{q^{\frac{2n-1}{2}}}{1+q^{2n-1}} 
\cos \left((2n-1) (z-\frac{\pi \tau}{2}) \right)\right] \,, \label{C8}
\end{eqnarray}
where we have used the expansion
\begin{equation*}
  \frac{1}{\sin z}= e^{iz}\frac{-2i}{1-e^{2iz}}= -2i\sum_{n=1}^\infty
  e^{i(2n-1)z} \,.
\end{equation*}
It is then manifest that $\Sigma$ does not have any other zeros in the
fundamental cell.

Given \eqref{C7} it is now also obvious that both sides of
\eqref{eq:26} have a single pole at $z=0$, and that the residue
is the same. Thus we have shown that both sides of \eqref{eq:26} have
the same poles and zeros (as a function of $z$).

Finally, it remains to prove that both sides of \eqref{eq:26} have the
same periodic properties in $z$. The \vth-functions are doubly
periodic with periods $\pi$ and $\pi + \pi\tau$ (see
appendix \ref{sec:theta}). From \eqref{eq:26} it is manifest that the 
right hand side changes sign under $z\mapsto z+ \pi$, as well as 
under $z\mapsto -z$. The behaviour of the left hand side is
identical, as can be seen from the periodicities of the
\vth-functions, as well as by direct inspection.  Under the 
shift  $z\mapsto z + \pi\tau$ the left hand side changes sign. For
the right hand side we observe that the bracket in $\Sigma$ 
\eqref{C8} transforms under $z\mapsto z + \pi\tau$ in the same way
as under $z\mapsto -z$. Since the factor of $\sin(z)$ in $\Sigma$
\eqref{C8} is cancelled in \eqref{eq:26} it then follows that the
right hand side therefore also changes sign under 
$z\mapsto z + \pi\tau$. Thus we have shown that the two sides of
\eqref{eq:26} have the same periodicity properties. Since their poles
and zeros agree, it then follows that they are indeed the same
function.

\providecommand{\bysame}{\leavevmode\hbox to3em{\hrulefill}\thinspace}

\end{document}